\documentstyle[prl,aps,multicol]{revtex}

\newcommand{\lsim}{\mathrel{\mathop{\kern 0pt \rlap
  {\raise.2ex\hbox{$<$}}}
  \lower.9ex\hbox{\kern-.190em $\sim$}}}
\newcommand{\gsim}{\mathrel{\mathop{\kern 0pt \rlap
  {\raise.2ex\hbox{$>$}}}
  \lower.9ex\hbox{\kern-.190em $\sim$}}}

\title{Quantum marking and quantum erasure for neutral kaons}

\author{A. Bramon$^{1}$, G. Garbarino$^{2}$ and B. Hiesmayr$^{1}$}

\address{$^1$Grup de F{\'\i}sica Te\`orica,
Universitat Aut\`onoma de Barcelona, E--08193 Bellaterra, Spain}
\address{$^2$Departament d'Estructura i Constituents de la Mat\`{e}ria,
Universitat de Barcelona, E--08028 Barcelona, Spain} \date{\today}

\begin{document}
\draft
\maketitle

\begin{abstract}
Entangled $K^0 \bar{K^0}$ pairs are shown to be suitable to
discuss extensions and tests of Bohr's complementarity principle through the
quantum marking and quantum erasure techniques suggested
by M. O. Scully and K. Dr$\ddot{\rm u}$hl [Phys. Rev. {\bf A 25}, 2208 (1982)].
Strangeness oscillations play the role of
the traditional interference pattern linked to wave--like behaviour, whereas the
distinct propagation in free space of the $K_S$ and $K_L$ components mimics
the two possible interferometric paths taken by particle--like objects.
\end{abstract}

%

\begin{multicols}{2}


Some twenty years ago Scully and Dr$\ddot{\rm u}$hl \cite{scully82} discussed
an interesting \emph{gedanken} experiment concerning the possibility to erase
information contained in quantum states and the effects that this
erasure can have on measurement outcomes. Further refinements of
the original proposal have appeared in recent years 
(see, for instance, Refs.~\cite{scully91,kwiat2}) 
and its connection with a 
central feature of quantum mechanics ---Bohr's complementarity 
principle--- has been widely debated. The quantum mechanical predictions,
including quantitative descriptions of complementarity \cite{englert}, have
been confirmed by a variety of recent interferometric experiments with    
atoms \cite{Durr} or photon pairs 
\cite{mandel90,mandel91,mandel91b,zeilinger95,kwiat,bjork,walborn,kim,bjork1}.

In this type of analyses one considers variations on the basic
double--slit experiment and exploits the well known complementarity
between the observation of interference fringes (wave--like behaviour) and
the acquisition of ``which way'' information (particle--like behaviour).
Interference patterns are observed if and only if in the two--way experiment
it is impossible to know, \emph{even in principle}, which way the particle took.
Interference disappears if there is a way to know ---e.g., through a  
\emph{quantum marking} procedure--- which
way the particle took; whether or not the outcome of the corresponding ``which way'' 
observation is actually read out, it does not matter: interference is in any way lost.

In most of the experiments performed up to date 
\cite{mandel90,mandel91,mandel91b,zeilinger95,kwiat,bjork,walborn,kim,bjork1} one 
uses two--photon entangled states from spontaneous parametric
down--conversion (SPDC). If the path of one photon is marked, 
information on the path taken by its entangled partner is in principle available
and no interference fringes  can be observed. But,
if that ``which way'' mark (usually, a specific polarization) is erased by means of 
a suitable measurement ---\emph{quantum erasure}---, observation of interference fringes
becomes possible in joint detection events.

The purpose of this Letter is to extend these considerations to
entangled pairs of neutral kaons. A neutral kaon beam presents the well known
phenomenon of $K^0$--$\bar{K^0}$ oscillations (in time), which will
play the role of the photon interference fringes (in  space).
Similarly, the short-- and long--lived kaon states, $K_S$ and $K_L$,
showing a distinct propagation in free--space, are the analogs 
of the two separated photon trajectories in interferometric devices. 
New forms of quantum markers and
erasers are thus offered and Bohr's complementary principle
extends its applicability. The CPLEAR experiment \cite{CPLEAR},
with $K^0 \bar{K^0}$ pairs created in $p\bar{p}$ annihilations at rest,
can be interpreted as a preliminary quantum eraser experiment.
The presently operating DA$\Phi$NE $\phi$--factory \cite{handbook} offers the
possibility to perform a complete test of complementarity 
by investigating both quantum marking and erasure  
along the lines we discuss in this note.


Two alternative bases, each one associated to a possible measurement on a neutral
kaon, have to be considered \cite{bg1}. The strangeness basis,
$\{K^0,\bar{K}^0\}$ with $\langle K^0|\bar{K}^0\rangle =0$,
is the appropriate one to discuss strong production and reactions of 
kaons, as well as kaon strangeness measurements.
Indeed, if a dense piece of ordinary (nucleonic) matter is
inserted along the neutral kaon trajectory, the incoming state is projected 
\emph{either} into $K^0$ (by strangeness conserving strong reactions such as $K^0 p \to K^+ n$)
\emph{or} into $\bar{K}^0$ ($\bar{K}^0 p \to \Lambda \pi^+$, $\bar K^0 n \to \Lambda \pi^0$,
$\bar{K}^0 n \to K^- p$). Nucleonic matter  plays here the same role as
a two--channel analyzer for polarized photon beams.

The second basis consists of the $K_S$ and $K_L$ states having well
defined masses $m_{S(L)}$ and decay widths $\Gamma_{S(L)}$; it is
the appropriate one to discuss neutral kaon propagation in free space,
with:
\begin{equation}
\label{evol}
|K_{S(L)}(\tau)\rangle = e^{-i\lambda_{S(L)}\tau} |K_{S(L)}\rangle ,
\end{equation}
and $\lambda_{S(L)}=m_{S(L)}-i\Gamma_{S(L)}/2$. The $K_S$ and $K_L$ eigenstates
preserve their own identity with (proper) time $\tau$, but,
since $\Gamma_S \simeq 579 \Gamma_L$, the short--lived component of a 
given neutral kaon extincts much faster than its long--lived one. 
Knowing if 
this kaon has propagated in free--space \emph{either} as  $K_S$ 
\emph{or}  $K_L$ is thus possible by detecting at which time it decays. If kaons
decaying before $\simeq 4.8\, \tau_{S}$ after production are identified as $K_S$'s and
those surviving after this time as $K_L$'s, misidentifications 
amount to only to a few parts in $10^{-3}$ \cite{bg1,eberhard}. 
Such a discrimination between the $K_S$ and $K_L$ propagation modes, 
linked to the obtention of ``\emph{which width}'' information, is thus in analogy 
with usual ``which way'' discussions in photon experiments.

The relationship between these two kaon bases is well known. Although one strictly has
$\langle K_{S} | K_{L} \rangle = 2({\rm Re}\, \epsilon) /(1 + |\epsilon |^2)
\simeq 3.2 \times 10^{-3}$, one can neglect this small CP violating
effect \cite{kabir} because it is of the same order as the previously mentioned
$K_S$ vs $K_L$  misidentifications and escapes from our control. With usual conventions
\cite{kabir}, one can then approximate:
\begin{eqnarray}
\label{basis}
|K^0\rangle &=&\left[|K_S\rangle + |K_L\rangle \right]/\sqrt{2} , \\
|\bar{K}^0\rangle &=&\left[|K_S\rangle - |K_L\rangle \right]/\sqrt{2}, \nonumber
\end{eqnarray}
and the two bases are associated with two (in a sense, maximally)
incompatible observables. Indeed, the proposed strangeness measurements and lifetime
observations completely exclude each other: the former require the
insertion of nucleonic matter, the latter propagation in free--space.
Bohr's complementarity principle is thus at work: if strangeness
(lifetime) is known, both  outcomes for lifetime (strangeness) are equally
probable, as follows immediately from  Eqs.~(\ref{basis}).


Suppose that at time $\tau=0$ a $K^0$ is produced via the strong reaction
$\pi^- p \to K^0 \Lambda$. From Eqs.~(\ref{evol}) and (\ref{basis}), at proper
time $\tau$ one has the following transition probabilities ($\Delta m 
\equiv m_L-m_S$):
\begin{eqnarray}
\label{s1}
|\langle K^0|K^0(\tau)\rangle|^2 &=&\left\{e^{-\Gamma_{S}\tau}+e^{-\Gamma_{L}\tau}\right\} \\
&&\times \left\{1+{\cal V}_0(\tau)\, {\rm cos}(\Delta m\, \tau)\right\}/4 ,  \nonumber \\
\label{s2}
|\langle \bar{K}^0|K^0(\tau)\rangle|^2 &=
&\left\{e^{-\Gamma_{S}\tau}+e^{-\Gamma_{L}\tau}\right\} \\
&&\times \left\{1-{\cal V}_0(\tau)\, {\rm cos}(\Delta m\, \tau) \right\}/4 , \nonumber \\
\label{l1}
|\langle K_L|K^0(\tau)\rangle|^2 &=& |\langle 
K_L|K_{L}(\tau)\rangle|^2 /2 = e^{-\Gamma_{L}\tau}/2 , \\
\label{l2}
|\langle K_S|K^0(\tau)\rangle|^2 &=& |\langle 
K_S|K_{S}(\tau)\rangle|^2 /2 =  e^{-\Gamma_{S}\tau}/2 .
\end{eqnarray}
Eqs.~(\ref{s1}) and (\ref{s2}) show the well known strangeness
oscillation phenomena with a  time dependent ``fringe visibility" 
${\cal V}_0(\tau)=1/\cosh(\Delta\Gamma \tau/2)$,
where $\Delta\Gamma \equiv \Gamma_L-\Gamma_S$.
On the contrary, Eqs.~(\ref{l1}) and (\ref{l2}) show that no $K_S$--$K_L$ 
oscillations are expected.

These observations admit the following interpretation. As soon as a  $K^0$ is
produced, it starts propagating in free space in the coherent superposition of
$K_S$ and $K_L$ given by Eqs.~(\ref{basis}) and mimics the two--way propagation of any 
system beyond a symmetrical double--slit. In the familiar double--slit case, the
system follows the two paths without ``jumping'' from one to the other, in
the same way as $K_S$--$K_L$ oscillations are ``forbidden''.
In the kaon case, however, there are not two separated trajectories but a single path 
comprising \emph{automatically} (i.e., with no need of any double--slit like apparatus) 
the two differently propagating components $K_S$ and $K_L$. At $\tau =0$ there is no 
information on which component actually propagates 
[``predictability'' of the path \cite{englert}, ${\cal P}(\tau=0)=0$] and 
the visibility of strangeness oscillations is
maximal, ${\cal V}_0(\tau=0) =1$. 
However, since the $K_S$ and $K_L$ components are intrinsically 
``marked'' by their different lifetimes, ``which width'' information is obtained for 
initial $K^0$'s  surviving up to $\tau$ and the corresponding oscillation visibility is 
now reduced. One easily finds 
${\cal P}(\tau ) = \left|\tanh(\Delta\Gamma \tau /2)\right|$, thus 
fulfilling the quantitative duality requirement
${\cal V}^2_0(\tau ) + {\cal P}^2(\tau ) =1$ \cite{englert} 
and supporting our present interpretation. 
 

Working with entangled kaon pairs, in the same way as one uses entangled pairs of
photons in analogous optical experiments, one can perform quantum 
marking and erasure.
To this aim, consider the decay of the $\phi(1020)$--meson resonance \cite{handbook}
(or, alternatively, $S$--wave $p\bar p$ annihilation \cite{CPLEAR}) into
$K^0 \bar{K}^0$ pairs. Just after the decay ($\tau=0$)
one has the following maximally entangled state:
\begin{eqnarray}
\label{entangled}
|\phi(0)\rangle & = & \frac{1}{\sqrt 2}\left[
|K^0\rangle_l |\bar{K}^0\rangle_r - |\bar{K}^0\rangle_l |K^0\rangle_r\right] \nonumber \\
& = & \frac{1}{\sqrt 2}\left[
|K_L\rangle_l |K_S\rangle_r - |K_S\rangle_l |K_L\rangle_r\right] , \nonumber
\end{eqnarray}
where $l$ and $r$ denote the ``left'' and ``right'' kaon directions of motion.
In the lifetime basis, the state evolution up to 
time $\tau_l$ ($\tau_r$) along the left (right) beam is given by:
\begin{eqnarray}
\label{notnorm}
&&|\phi(\tau_l,\tau_r)\rangle = \frac{1}{\sqrt 2}\left\{
e^{-i(\lambda_L\tau_l+\lambda_S\tau_r)}|K_L\rangle_l|K_S\rangle_r \right. \nonumber \\
&&\hphantom{|\phi(\Delta\tau)\rangle = \frac{N(\tau_l,\tau_r)}{\sqrt 2}}
\left. -e^{-i(\lambda_S\tau_l+
\lambda_L\tau_r)}|K_S\rangle_l|K_L\rangle_r\right\} . \nonumber
\end{eqnarray}

It will be convenient to consider only kaon pairs with both left and right members surviving up to 
$\tau_{l}$ and $\tau_{r}$. These are described by the following normalized state:
\begin{eqnarray}
\label{time}
|\phi(\Delta\tau)\rangle&=& \frac{1}{\sqrt {1+e^{\Delta\Gamma \Delta\tau}}}
\biggl\{|K_L\rangle_l|K_S\rangle_r  \\
&& \qquad\qquad -e^{i \Delta m \Delta\tau} e^{{1 \over 2} \Delta \Gamma \Delta\tau}
|K_S\rangle_l|K_L\rangle_r\biggl\} ,\nonumber
\end{eqnarray}
where $\Delta\tau\equiv \tau_l-\tau_r$. Eq. (\ref{time}) can be rewritten as 
\begin{eqnarray}
\label{timestrangeness}
&&|\phi(\Delta\tau)\rangle =
\frac{1}{\sqrt {2(1+e^{\Delta\Gamma \Delta\tau})}} 
\biggl\{|K^0 \rangle_l|K_S\rangle_r -|\bar K^0 \rangle_l|K_S\rangle_r \nonumber \\
&& \quad -e^{i \Delta m \Delta\tau} e^{{1 \over 2} \Delta \Gamma \Delta\tau}
\left[|K^0 \rangle_l|K_L\rangle_r + |\bar K^0 
\rangle_l|K_L\rangle_r\right]\biggl\} ,
\end{eqnarray}
or, in the strangeness basis, as
\begin{eqnarray}
\label{bothstrangeness}
&&|\phi(\Delta\tau)\rangle = \frac{1}{2 \sqrt {1+e^{\Delta\Gamma \Delta\tau}}} \\
&&\times \left\{(1-e^{i \Delta m \Delta\tau} e^{{1 \over 2} \Delta \Gamma \Delta\tau})
\left[|K^0\rangle_l|K^0\rangle_r-|\bar K^0\rangle_l|\bar K^0\rangle_r\right] \right.  \nonumber \\
&& \left. +(1+e^{i \Delta m \Delta\tau} e^{{1 \over 2} \Delta \Gamma 
\Delta \tau})
\left[|K^0\rangle_l|\bar K^0\rangle_r-|\bar
K^0\rangle_l|K^0\rangle_r\right] \right\} .\nonumber
\end{eqnarray} 
Eqs.~(\ref{time})--(\ref{bothstrangeness}) 
immediately supply the various joint 
probabilities $P(K_{l},K_{r})$ for detecting a $K_{l}$ ($K_{r}$) on the left 
(right) at time $\tau_{l}$ ($\tau_{r}$) 
with $K_{l,r} = K^0, \bar K^0 , K_{S}$ or $K_{L}$.

By normalizing to
surviving kaon pairs, one does not need to consider decay--product states.
Thanks to this, the entangled states (\ref{time})--(\ref{bothstrangeness}) 
only depend on the time difference $\Delta\tau$  
and we can work with bipartite two--level quantum systems as in the 
optical case. For details concerning an accurate description 
of the kaon pair evolution in $\tau_l$, $\tau_r$
see Refs.~\cite{ghirardi,BH1}.   

Quantum marking and erasure for these entangled neutral kaon 
pairs can be described as follows. One always measures 
the strangeness of the left moving kaon; by making this measurement at different values of 
$\tau_l$ one can look for strangeness oscillations of this  
kaon ---the \emph{signal} or \emph{object} kaon. 
Conversely, the measurement on the right moving ---\emph{idler} or 
\emph{meter}--- kaon is always performed at a \emph{fixed} time 
$\tau^0_r$; but, by inserting or not a strangeness detector   
in two alternative set--ups, either strangeness or lifetime will be 
measured at $\tau^0_r$. In this way, various $\tau_l$--dependent 
probabilities for object--meter joint detections are recorded. 

Without the strangeness detector on the right beam, 
one can observe the decay of the freely propagating meter kaon, which 
will be identified either as $K_{S}$ or $K_{L}$ \cite{foot1}.
The acquisition of 
this ``which width'' information implies the 
corresponding one for the object kaon and therefore strangeness 
oscillations 
(in $\Delta\tau\equiv \tau_l-\tau_r^0$)  
should not be visible for any of the four possible joint detection probabilities 
$P(K_{l},K_{r})$ with $K_{l} = K^0$ or $\bar K^0$ and  $K_{r} = K_{S}$ 
or $K_{L}$. This is immediately seen from Eq.~(\ref{timestrangeness}). 

However, the possibility to obtain ``which width'' information can be 
prevented by quantum erasure, i.e., by measuring strangeness on 
the meter kaon at time $\tau^0_r$. 
The four joint probabilities $P(K_{l},K_{r})$
with $K_{l,r} = K^0$ or $\bar K^0$
will then show the $\tau_l$--dependent 
strangeness oscillations and anti--oscillations immediately 
deducible from Eq.~(\ref{bothstrangeness}).
One has then revived the same $K_S$--$K_L$ interference effects 
typical of the single kaon case [Eqs.~(\ref{s1}), (\ref{s2})], 
with visibility  ${\cal V}(\tau_l)=1/{\rm cosh}[\Delta \Gamma 
(\tau_l-\tau^0_r)/2]$. The erasure is maximal 
(${\cal V}=1$) for $\tau_l=\tau^0_r$ (no ``which width'' information
available), but the
strangeness oscillations disappear  totally (${\cal V}\to 0$) 
for $\tau_l-\tau^0_r \to \infty$ (full ``which width'' information available).  
The time--dependent ``distinguishability'' of the path \cite{englert},
quantifying ``which width'' information, is given by 
${\cal D}(\tau_l)=\left|\tanh[\Delta \Gamma (\tau_l-\tau^0_r)/2]\right|$
and satisfies the \emph{duality relation} 
${\cal V}^2(\tau_l)+{\cal D}^2(\tau_l)=1$ \cite{englert}.


In a series of optical experiments, entangled pairs of photons 
have been produced in states similar to that in Eq.~(\ref{time}).
In one of the experiments of 
Ref.~\cite{zeilinger95}, two interfering two--photon amplitudes 
are prepared by forcing a pump beam to cross \emph{twice} the same 
non--linear crystal. Idler and signal photons from the first 
down--conversion are ``marked'' by rotating their polarization by 
$90^\circ$ and then superposed to the idler and signal photons 
emerging later from the second passage of the pump beam through the crystal. 
If type--II SPDC were used, one would obtain the entangled state:
\begin{equation}
\label{state_herzog}
|\Psi \rangle = \left[ |V\rangle_i |H\rangle_s -
e^{i \Delta \phi}|H\rangle_i |V\rangle_s \right]/\sqrt{2} ,
\end{equation}
where the term $|V\rangle_i |H\rangle_s$ ($|H\rangle_i |V\rangle_s$) 
refers to photon pairs 
produced in the second (first) passage and the relative phase $\Delta\phi$ is under 
control by the experimenter. 
Thanks to entanglement, the distinct vertical or horizontal idler
polarizations supply full ``which path'' information for the signal photons 
and no interference can be observed in signal--idler joint detections. 
To erase this information, the idler (signal) photons 
have to be detected after crossing polarization analyzers placed at $45^\circ$ 
or $-45^\circ$ ($45^\circ$), i.e., the idler and signal polarizations 
must be jointly measured in a basis consisting of the two 
symmetric and antisymmetric superpositions 
$|H\rangle \pm |V\rangle$ as in Eqs.~(\ref{basis}) \cite{note}.
Similarly, the photonic experiments of Refs.~\cite{mandel90,mandel91} use 
two SPDC processes and 
could be discussed along the same lines, but their interpretation 
as quantum erasers is less clear \cite{kwiat2,mandel91b,Ou97}.

The similarity between the (``which way'' marked) two--photon state 
(\ref{state_herzog}) and the (``which width''
marked) two--kaon state (\ref{time}) is obvious. The latter, however, is
automatically prepared in $\phi$ decays or $S$--wave $p\bar{p}$ 
annihilations. These processes can be simply viewed as coherently producing 
a superposition of the $|K_L\rangle_l |K_S\rangle_r$ and $|K_S\rangle_l |K_L\rangle_r$
two--kaon amplitudes. 
The kaon mass difference $\Delta m$ introduces automatically a time 
dependent relative phase between the two amplitudes, as it is evident
from Eq.~(\ref{time}). The 
marking and erasure operations can be performed on entangled kaon pairs 
as in the optical case previously discussed: 
photon detection after a polarization analyzer at 
$\theta=45^\circ$ ($-45^\circ$) corresponds to $K^0$ ($\bar{K}^0$) detection, 
while $\theta=0^\circ$ ($90^\circ$) corresponds to $K_S$ ($K_L$) 
observation. The additional exponential decay factor, 
$e^{\frac{1}{2}\Delta \Gamma (\tau_l -\tau^0_{r})}$, 
which is obviously absent for stable photons, 
reduces the fringe  visibility in kaon experiments.


Another series of photon experiments \cite{kwiat,bjork,walborn}
followed a different strategy. Instead of using two 
SPDC processes, a pump beam is  
sent to a \emph{single} non--linear crystal. Each member of the emerging 
down--converted photon pair is then directed towards a 50\%--50\% beam--splitter 
$BS$ in the set--ups of Refs.~\cite{kwiat,bjork}. 
In Ref.~\cite{kwiat}, a half--wave plate is inserted along
one of the two paths before crossing the $BS$
in order to introduce path distinguishability.   
In Ref.~\cite{bjork}, the two beams emerging from that first $BS$ 
are sent towards two well separated beam--splitters, thus producing two 
amplitudes marked by the position of the beam--splitter they come from.
In the experiment of Ref.~\cite{walborn}, one of the two 
down--converted photons is directed towards a double--slit:  
two appropriately placed quarter--wave plates 
mark each path beyond the
double--slit with left or right circular polarizations.
In all these experiments \cite{kwiat,bjork,walborn}, 
a two--photon entangled state similar to that in 
Eq.~(\ref{state_herzog}) is obtained after 
properly postselecting the state directly produced by the interferometers. 
If these (position or polarization) marks are not erased, interference 
effects are never seen; by erasing with appropriate measurements,  
which project on symmetric and antisymmetric states as in Eqs.~(\ref{basis}),
interferences reappear. 

Since the $\phi$ resonance essentially consists of a strange--antistrange 
$s \bar{s}$ quark pair, the decay $\phi \to K^0 \bar{K}^0$ 
proceeds by pumping a $d \bar{d}$ pair from the 
vacuum, which then recombine into a $K^0(d \bar{s})$--$\bar K^0(s \bar{d})$ meson pair. 
This decay process, like SPDC from a \emph{single} non--linear crystal, 
produces good quality two--particle entangled states. 
Indeed, each $K^0 \bar K^0$ pair 
produced in a $\phi$ decay automatically propagates in free space as the 
coherent superposition of $|K_{L}\rangle_l |K_{S}\rangle_r$ and 
$|K_{S}\rangle_l |K_{L}\rangle_r$ previously discussed. 
The optical elements (beam--splitters or double slits) needed in the 
experiments of Refs.~\cite{kwiat,bjork,walborn} to produce 
potentially interfering two--photon
amplitudes are automatically provided by Nature for neutral kaons
[see Eq.~(\ref{bothstrangeness})], with no need for postselection.


To summarize, we have explored Bohr's principle in a new
direction: the complementarity between strangeness and lifetime measurements
on neutral kaons (this closes the issue, since no other projective measurement 
can be performed). We would like to emphasize that 
the neutral kaon system reveals to be suitable
for an optimal demonstration \cite{scully91,kwiat2} of quantum erasure: (1) 
since ``which width'' information is carried by a system (the meter kaon) 
distinct and spatially separated from the
interfering system (the object kaon), the marking and erasure operations can 
be performed in the ``delayed choice'' mode ($\tau_l<\tau^0_r$); 
(2) single--particle states (as opposed to coherent photon states) are 
detected on each side; 
(3) the entanglement with the meter kaon, necessary to perform 
marking or erasure, does not disturb the
states $|K_S\rangle$ and $|K_L\rangle$ of the object kaon; 
(4) quantum erasure with entangled kaons allows one to restore the same 
interference phenomenon between the $K_S$ and $K_L$
propagation modes exhibited by a single kaon state 
produced as $K^0$ or $\bar{K}^0$. 

An experimental test of the marking and erasure operations we have discussed is highly
desirable and should be feasible at $\phi$--factories and $p\bar{p}$
machines. Actually, the CPLEAR collaboration \cite{CPLEAR} has already done part of the 
work required: the two experimental points (for $|\Delta \tau|$= 0 and 1.2$\tau_S$) collected
by this experiment reproduced the 
joint strangeness oscillations predicted by quantum
mechanics [see Eq.~(\ref{bothstrangeness})]. 
New measurements confirming with better precision these 
oscillations for a larger range of $\Delta \tau$ values, 
as well as the non--oscillating behaviour when ``which width'' information
is in principle available [see Eq.~(\ref{timestrangeness})], 
are needed for a full complementarity test.    

Work partly supported by EURIDICE
HPRN--CT--2002--00311, BFM--2002--02588, MA 7--1886/02 and INFN.

\end{multicols}


\begin{references}
\bibitem{foot1}
When $\Delta \tau = \tau_{l} - \tau_{r} =0$, each one of the two members of a surviving kaon 
pair has the same probability for being a $K_{S}$ or a $K_{L}$. 
For $\Delta \tau = \tau_{l} - \tau_{r} > 0$ ($<0$), 
the left (right) moving kaon is more probably a $K_{L}$ than its right 
(left) moving partner. With this additional information, a simple calculation
shows that one has to identify as right going $K_{S}$'s 
all kaons decaying between 
$\tau_{r}$ and $\tau_r +\Delta \tau_{r}$, 
where $\Delta \tau_r$ must satisfy 
$\Delta \Gamma \Delta \tau+\Gamma_S \Delta \tau_r+{\rm ln(1-e^{\Gamma_L \Delta \tau_r})}=0$;
undecayed kaons have to be identified as $K_{L}$'s. Then, the 
$K_S$ and $K_L$ misidentification probabilities coincide and reduce,
as in the single kaon case, to a few parts in $10^{-3}$.
\bibitem{note}
Note that the presence of the signal polarization analyzer
at $45^\circ$, which is essential to perform
quantum erasure, does not alter by itself the interference--free situation
obtained when the idler polarization analyzer is placed at
$0^\circ$ or $90^\circ$.
\end{references}
\end{document}